# Multimodal Visual Sensing of Temperature and Pressure: From Spectroscopic Readout to Multiple Linear Regression-Enhanced RGB Analysis

**Maja Szymczak[a,*], Yufan Meng[b], Guanjun Xiao[b,*], Miguel A. Hernández-Rodríguez[c], Iga Sawaryn[a], Bo Zou[b], Lukasz Marciniak[a,*]**

[a] Institute of Low Temperature and Structure Research Polish Academy of Sciences, 50-422 Wroclaw, Poland

[b] State Key Laboratory of High Pressure and Superhard Materials, College of Physics, Jilin University, Changchun 130012, China

[c] Departamento de Física, and IUdEA, Universidad de La Laguna Apdo. Correos 456, E-38200 San Cristóbal de La Laguna, Santa Cruz de Tenerife, Spain

*corresponding authors: m.szymczak@intibs.pl, xguanjun@jlu.edu.cn, l.marciniak@intibs.pl

## Abstract

Transforming a single luminescent host into distinct temperature- and pressure-sensing systems through composition control provides a powerful route toward multifunctional optical sensors. Here, we introduce a concentration-tunable $KGaGeO_4$:$Bi^{3+}$,$Eu^{3+}$ platform whose sensing function can be selectively directed toward thermometry or manometry by adjusting the dopant balance. Spectroscopically distinct $Bi^{3+}$ centers and $Eu^{3+}$ emission exhibit differentiated responses to temperature and pressure, enabling multimodal readout through ratiometric luminescence, chromaticity coordinates, spectral shifts, and visible color changes. Importantly, these color changes were translated into quantitative temperature and pressure maps using RGB imaging combined with multiple linear regression (MLR). By simultaneously exploiting multiple color channels, MLR increased the maximum relative thermal sensitivity from approximately 1% $K^{-1}$ for conventional RGB ratios to 8.8% $K^{-1}$ and the pressure sensitivity from 169% $GPa^{-1}$ to nearly 721% $GPa^{-1}$. This work presents the first application of MLR-assisted RGB analysis for quantitative luminescence pressure sensing. The proposed strategy integrates composition-controlled functionality, multimodal spectroscopic sensing, direct visual readout, and data-assisted imaging, demonstrating the synergy between material engineering and multivariate analysis for highly sensitive multifunctional optical sensing.

## 1. Introduction

In the era of increasing automation and digitalization of technological processes, the development of sensing platforms capable of reliable, sensitive, and straightforward readout of physical parameters has become increasingly important. Luminescence-based sensing is particularly attractive in this context because it combines high sensitivity with remote readout and, when appropriately designed, can enable the simultaneous determination of multiple physical parameters. Recent advances in luminescent sensing have demonstrated that different spectroscopic observables can be selectively assigned to different external stimuli, allowing temperature and pressure, among others, to be determined independently while minimizing cross-sensitivity between the corresponding readout channels.[1] A major advantage of luminescence-based sensing is the wide variety of available readout strategies. Depending on the application, sensing can be based on emission intensity, spectral position, bandwidth, luminescence kinetics, or ratios between selected spectroscopic parameters.[2] Time-resolved approaches are particularly advantageous in biological and optically complex environments because luminescence kinetics is, to a large extent, independent of phosphor concentration, excitation intensity, optical path length, and detector sensitivity.[3,4] In contrast, ratiometric sensing based on the luminescence intensity ratio (*LIR*) offers a particularly attractive solution for many technological applications because it provides an internally referenced signal while maintaining a comparatively simple and inexpensive detection scheme. The use of two spectrally distinguishable emission components additionally reduces the influence of fluctuations in excitation power, sample amount, and collection geometry, thereby improving the reliability of the measurement.[5,6] Beyond conventional point sensing, luminescent materials offer another important advantage: they can provide spatially resolved information over extended surfaces and, in suitable systems, within optically accessible volumes. This capability distinguishes luminescence thermometry from many conventional temperature-measurement techniques. Thermocouples, resistance temperature detectors, and thermistors provide only local temperature information and require the sensing element to be physically introduced into or attached to the specific region of interest, which can be challenging in confined, miniaturized, or geometrically complex systems and does not readily enable spatial temperature mapping. Infrared thermography enables non-contact two-dimensional temperature mapping; however, the measured signal generally corresponds to the temperature of the first accessible surface. Consequently, when the monitored component is covered by an optically transparent or semi-transparent layer, encapsulated within a polymer, protected by quartz, glass, or another dielectric material, the apparent surface temperature may differ from the actual temperature of the functional component because of thermal gradients and differences in thermal conductivity. Such limitations are especially relevant for miniaturized devices, encapsulated systems, coatings, and components in which direct placement of an electrical sensor is impractical. Luminescent thermometry provides an attractive alternative because the temperature-sensitive phosphor itself can be incorporated directly into, deposited onto, or positioned close to the region of interest. Spatially resolved temperature information can then be extracted from luminescence images rather than from individual point measurements. In its simplest implementation, an image of the emitting material is acquired by the camera and separated into the red, green, and blue color channels. Ratios between the RGB intensities can subsequently be correlated with temperature using an appropriate calibration function.[7,8] This approach is particularly appealing because it does not necessarily require a dedicated spectrometer or a complex and expensive detection system. Provided that the acquisition conditions are properly controlled, commercially available digital cameras can be employed as detectors, considerably reducing the complexity and cost of the measurement platform. Such RGB-based luminescence thermometry has already been successfully demonstrated for spatial temperature visualization and represents an important step toward simple, portable, and readily implementable optical sensing systems.[7–9] A conceptually similar imaging-based strategy can also be envisaged for luminescent pressure sensing, in which pressure-induced changes in emission color are translated into spatially resolved RGB parameters. However, color-image-based approaches remain considerably less explored in luminescence manometry than in thermometry.

The performance of color-based readout can be further enhanced by moving beyond simple ratios between individual RGB channels. In particular, multivariate approaches can exploit information contained simultaneously in several image parameters. Multiple linear regression (MLR), for example, enables several partially correlated optical descriptors to be combined into a single calibration model, potentially improving sensitivity, precision, and robustness compared with calibration based on a single intensity ratio.[10–12] Such approaches are especially attractive for luminescent materials exhibiting pronounced changes in spectral shape and emission color, because the information encoded in the complete RGB response can be utilized rather than reduced to one pair of channels. The integration of luminescent sensing with simple digital imaging and multivariate analysis therefore represents a promising route toward low-cost, spatially resolved, and highly sensitive optical sensing. Although multivariate regression has already been successfully implemented in luminescence thermometry and has also been introduced into luminescence manometry, its combination with RGB image analysis for pressure sensing has not yet been demonstrated.

In this work, $KGaGeO_4$ co-doped with $Bi^{3+}$ and $Eu^{3+}$ is demonstrated as a multifunctional luminescent platform for temperature and pressure sensing. The coexistence of broad blue-green $Bi^{3+}$ emission and narrow red $Eu^{3+}$ emission provides strong spectral and chromatic contrast, while variation of the $Eu^{3+}$ concentration enables the relative contribution of both emitters to be precisely tuned. As a consequence, the overall emission color can be systematically shifted from $Bi^{3+}$-dominated blue-green or turquoise emission toward increasingly red emission with higher $Eu^{3+}$ contribution. This pronounced color tunability is particularly advantageous for visual sensing, because changes in the relative intensities of the $Bi^{3+}$ and $Eu^{3+}$ bands can be directly translated into easily distinguishable chromatic responses. This compositional dependence was exploited to obtain two complementary sensing regimes: a higher-$Eu^{3+}$ composition exhibiting pronounced temperature-dependent spectral and color changes was selected for thermometry, whereas a low-$Eu^{3+}$ composition characterized by substantially weaker temperature dependence was employed for pressure sensing, thereby reducing temperature-induced interference in the manometric readout. Importantly, $Bi^{3+}$ and $Eu^{3+}$ exhibit markedly different responses to temperature and pressure, enabling the two stimuli to be distinguished through appropriately selected compositions and spectroscopic parameters. Beyond conventional spectroscopic readouts, including *LIR*, spectral shifts, and CIE1931 coordinates, the pronounced redistribution between the blue-green $Bi^{3+}$ and red $Eu^{3+}$ emission components was exploited for direct color-based sensing. RGB parameters extracted from luminescence images were used to establish temperature- and pressure-dependent calibration relationships, providing a simple imaging-based alternative to spectroscopic detection. Furthermore, MLR was combined with RGB image analysis to exploit the temperature- and pressure-dependent information encoded simultaneously in multiple color-channel ratios and to enhance the sensing performance beyond that achievable using individual RGB parameters. Notably this work represents the first demonstration of the synergistic use of RGB luminescence imaging and MLR for pressure sensing. The proposed approach therefore integrates compositional tuning, excitation selectivity, conventional luminescence sensing, RGB imaging, and multivariate analysis, providing independent temperature and pressure readout while minimizing cross-sensitivity between the two parameters and extending image-assisted multivariate sensing from luminescence thermometry toward manometry (Figure 1).

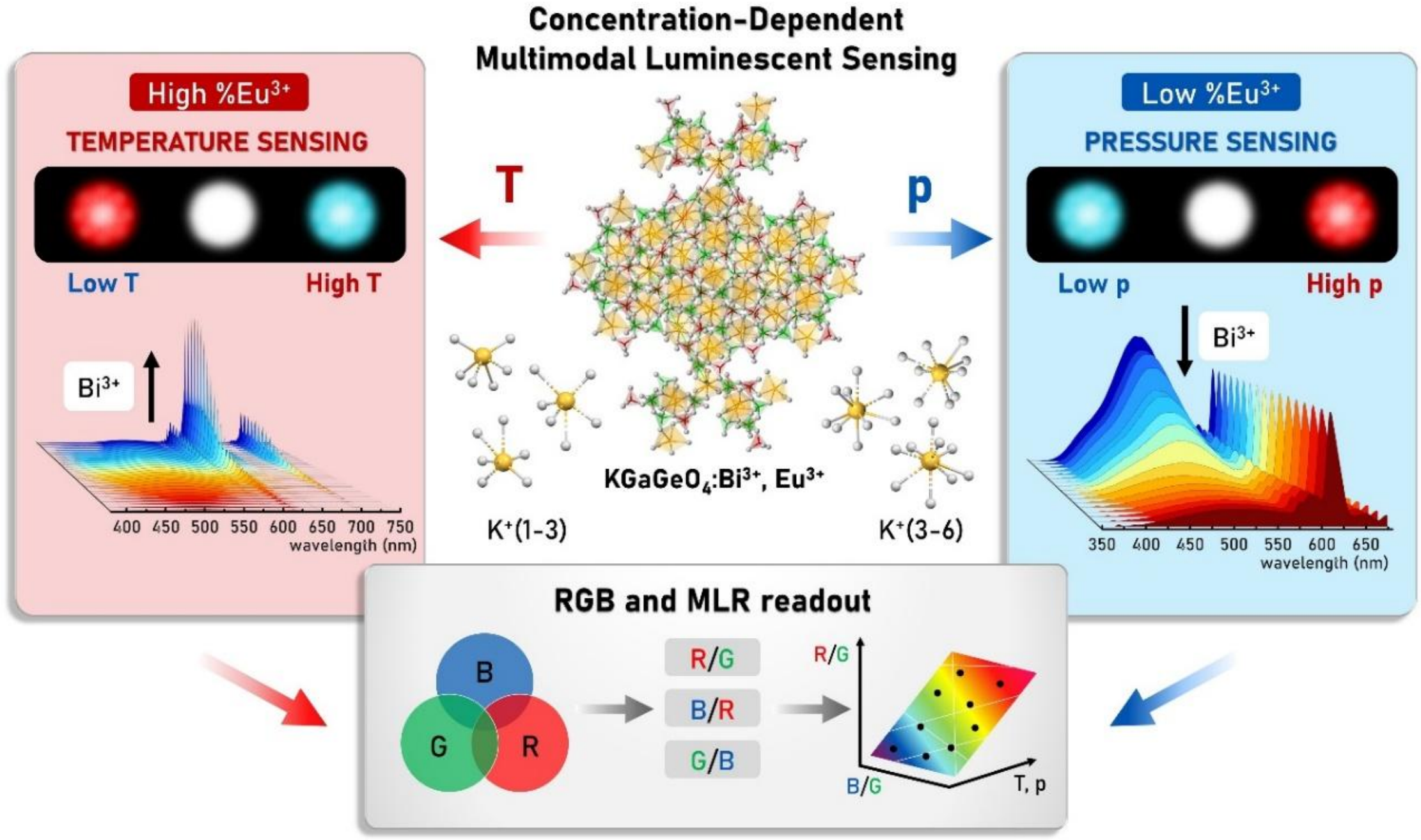

**Figure 1.** Conceptual diagram illustrating the concentration-dependent multifunctional luminescent sensing concept of $KGaGeO_4:Bi^{3+},Eu^{3+}$. Variation of the $Eu^{3+}$ concentration enables the phosphor response to be tailored toward either temperature or pressure sensing while minimizing cross-sensitivity between the two parameters. The distinct responses of the broad $Bi^{3+}$ emission and the narrow $Eu^{3+}$ emission to temperature and pressure result in pronounced stimulus-dependent color changes, which can be directly exploited through RGB-based optical readout. Analysis of then ratios of the color channels provide simple calibration parameters for temperature and pressure determination, while implementation of multiple linear regression (MLR) enables simultaneous use of multichannel color information to improve the sensitivity of the sensing readout.

## 2. Experimental Section

*Materials*

All precursors for the synthesis were used without additional purification: $K_2CO_3$ (Thermo Fisher Scientific, 99.997% of purity), $Ga_2O_3$ (Thermo Fisher Scientific, 99.999% of purity), $GeO_2$ (Thermo Fisher Scientific, 99.9999% of purity), $Bi_2O_3$ (Thermo Fisher Scientific, 99.999% of purity) and $Eu_2O_3$ (Thermo Fisher Scientific, 99.99% of purity).

*Synthesis*

A high-temperature solid-state reaction method was employed to synthesize undoped $KGaGeO_4$, $KGaGeO_4$:1%$Bi^{3+}$, and $KGaGeO_4$:1%$Bi^{3+}$, x%$Eu^{3+}$, where x = 0.1-5. The appropriate amounts of the starting reagents were weighed in stoichiometric proportions and thoroughly ground in an agate mortar using hexane as a dispersing medium to improve homogenization. The resulting mixtures were transferred to corundum crucibles and heated in air at 1323 K for 10 h with a heating rate of 10 K $min^{-1}$. After natural cooling to room temperature, the obtained powders were reground to ensure uniformity.

*Characterization*

Powder X-ray diffraction was employed to verify the structural characteristics and phase purity of the prepared materials. The measurements were carried out on a PANalytical X'Pert Pro diffractometer using Ni-filtered Cu Kα radiation at an operating voltage of 40 kV and a current of 30 mA. The morphology of the powders was investigated by scanning electron microscopy using an FEI Nova NanoSEM 230 microscope. Elemental mapping of selected particles was performed by means of an EDAX Genesis XM4 energy-dispersive X-ray spectroscopy system coupled to the microscope. For sample preparation, the powders were suspended in methanol, deposited dropwise onto carbon-coated aluminum stubs, and dried under an infrared lamp prior to imaging. Raman spectra were recorded using an RMS-1000 Raman microscope (Edinburgh Instruments) equipped with a 532 nm excitation laser. The measurements were performed using a 20x microscope objective and an 1800 lines $mm^{-1}$ diffraction grating. Luminescence measurements (also as a function of temperature) were collected using an FLS1000 fluorescence spectrometer (Edinburgh Instruments). Excitation spectra were recorded using a 450 W xenon arc lamp, whereas decay measurements were performed under excitation with a 60 W xenon flash lamp.

The luminescence decay profiles were described using a biexponential function:

$$\tau_{avr} = \frac{A_1\tau_1^2 + A_2\tau_2^2}{A_1\tau_1 + A_2\tau_2} \quad \textbf{(1)}$$

$$I(t) = I_0 + A_1 \cdot \exp\left(-\frac{t}{\tau_1}\right) + A_2 \cdot \exp\left(-\frac{t}{\tau_2}\right) \quad \textbf{(2)}$$

where $\tau_1$ and $\tau_2$ are the individual lifetime components, and $A_1$ and $A_2$ correspond to their respective pre-exponential amplitudes in the biexponential fit.

Temperature-dependent experiments were conducted using a Linkam THMS 600 heating-cooling stage. The temperature stability and set-point resolution of the system were both 0.1 K. At each temperature point, a stabilization time of 2 min was applied before the spectroscopic data were collected.

Pressure-dependent luminescence measurements were performed using a screw-driven diamond anvil cell (DAC). The powdered sample was loaded into a 150 μm diameter hole drilled in a T301 stainless-steel gasket pre-indented to a thickness of approximately 45 μm. Silicone oil (10 cSt, Aladdin) was used as the pressure-transmitting medium. The pressure inside the DAC was determined from the pressure-induced shift of the ruby fluorescence lines. High-pressure PL spectra were recorded with an optical fiber spectrometer (Ocean Optics, QE65000). A semiconductor laser with an excitation wavelength of 355 nm was employed for all fluorescence experiments. The laser beam controlled by an adjustable attenuator was focused on the sample with a 20-μm spot 20× UV Plan apochromatic objective and passed through a filter to eliminate effects on acquisition.

Digital luminescence images were acquired using a Canon EOS 400D camera with an exposure time of 1 s. For the proof-of-concept experiment, the recorded images were separated into their individual red (R), green (G), and blue (B) channels. The corresponding channel intensities were subsequently used to calculate intensity-ratio maps, enabling spatial visualization of the luminescence response. Image processing and RGB channel extraction were performed using IrfanView 64 (version 4.51).

## 3. Results and discussion

$KGaGeO_4$ belongs to the stuffed-tridymite family of ($ABCO_4$)-type compounds and crystallizes in a hexagonal structure with the *P6₃* space group.[13,14] Its framework is built primarily from corner-connected $GaO_4$ and $GeO_4$ tetrahedra, as illustrated in Figure 2a. The channels and cavities formed by this tetrahedral framework are occupied by $K^+$ ions. Importantly, the crystal structure contains six crystallographically nonequivalent $K^+$ sites, which can be divided into two groups according to their local coordination environment: three six-coordinated and three nine-coordinated sites (Figure 2b). The presence of several nonequivalent $K^+$ positions is particularly relevant in the context of $Bi^{3+}$ and $Eu^{3+}$ incorporation. Although substitution of trivalent ions at monovalent $K^+$ sites introduces a local charge mismatch, occupation of the $K^+$ sublattice appears more plausible than substitution at the considerably smaller tetrahedrally coordinated $Ga^{3+}$ or $Ge^{4+}$ sites. The latter are characterized by substantially smaller coordination polyhedra, making accommodation of the relatively large $Bi^{3+}$ and $Eu^{3+}$ ions energetically less favorable. Substitution at $K^+$ positions would, however, necessarily be accompanied by appropriate charge-compensating defects, such as cation vacancies and/or other local structural rearrangements. Since the local coordination geometry may strongly influence the electronic structure and spectroscopic response of incorporated luminescent ions, the individual $K^+$-$(O^{2-})_x$ polyhedra were analyzed in more detail. For each crystallographically independent $K^+$ site, the $K^+$-$O^{2-}$ distances were determined and the degree of polyhedral distortion was quantified using the distortion index ($D$)[15,16]:

$$D = \frac{1}{n}\sum_{i=1}^{n}\frac{|d_i - d_{av}|}{d_{av}} \quad \textbf{(3)}$$

where $d_i$ represents the individual $K^+$-$O^{2-}$ distance, $d_{av}$ is the average $K^+$-$O^{2-}$ distance within a given coordination polyhedron, and $n$ is the coordination number. The resulting structural parameters are summarized in Table 1, below.

**Table 1.** Structural parameters of the individual $K^+$-$(O^{2-})_x$ coordination polyhedra in $KGaGeO_4$.

| ***SITE*** | ***CN*** | ***$d_{avr}$ (Å)*** | ***D*** | ***D (%)*** |
|---|---|---|---|---|
| ***$K^+$(1)*** | 6 | 2.894 | 0.0989 | 9.89 |
| ***$K^+$(2)*** | 6 | 2.900 | 0.0918 | 9.18 |
| ***$K^+$(3)*** | 6 | 2.877 | 0.0804 | 8.04 |
| ***$K^+$(4)*** | 9 | 2.994 | 0.0432 | 4.32 |
| ***$K^+$(5)*** | 9 | 3.008 | 0.0417 | 4.17 |
| ***$K^+$(6)*** | 9 | 2.993 | 0.0419 | 4.19 |

A clear difference is observed between the two groups of $K^+$ sites. The six-fold coordinated $K^+$(1-3) environments exhibit substantially larger distortion indices than the nine-fold coordinated $K^+$(4-6) sites, with $D$ values approximately twice as high. Thus, the $K^+$(1-3) sites represent considerably more asymmetric and distorted local environments, whereas $K^+$(4-6) are characterized by more regular coordination. This distinction may be particularly important for understanding the preferential incorporation of $Bi^{3+}$ and $Eu^{3+}$ ions. The $6s^2$ electronic configuration of $Bi^{3+}$ is associated with a stereochemically active lone pair, making its optical properties highly sensitive to local asymmetry and structural relaxation. Consequently, the more distorted six-fold coordinated $K^+$ sites may provide favorable environments for $Bi^{3+}$ incorporation and may contribute to the formation of strongly perturbed $Bi^{3+}$ luminescent centers.[17,18] In contrast, $Eu^{3+}$ commonly adopts higher coordination numbers, typically ranging from 8 to 9, suggesting that the nine-fold coordinated $K^{+}$(4-6) environments may be structurally more suitable for $Eu^{3+}$ accommodation.[19] On this basis, preferential occupation of the more distorted $K^{+}$(1-3) sites by $Bi^{3+}$ and of the higher-coordination $K^{+}$(4-6) sites by $Eu^{3+}$ can be proposed as a plausible structural model. Nevertheless, this assignment should be regarded as a working hypothesis rather than direct evidence of site occupation and will therefore be further discussed in relation to the spectroscopic properties of the investigated materials.

X-ray diffraction analysis confirmed that the investigated $Bi^{3+}/Eu^{3+}$-codoped materials retained the $KGaGeO_4$ crystal structure without detectable secondary phases (Figure 2c). Additional information on the local vibrational structure was obtained from the Raman spectrum of undoped $KGaGeO_4$ recorded at 93 K (Figure 2d). Three main spectral regions can be distinguished. The low-wavenumber region, approximately 300-350 $cm^{-1}$, is dominated by lattice vibrations and low-frequency motions of the tetrahedral framework. Bands located between approximately 350 and 600 $cm^{-1}$ are predominantly associated with bending and deformation modes of the $GaO_4$ and $GeO_4$ tetrahedra nd the corresponding X-O-X linkages (X = Ga, Ge. In the higher-wavenumber region, between approximately 750 and 900 $cm^{-1}$, the Raman bands originate mainly from stretching vibrations of the tetrahedral framework, with a particularly significant contribution from Ge-O stretching modes. The overall distribution of the Raman bands is consistent with that expected for tetrahedrally coordinated $GaO_4/GeO_4$-based frameworks and further supports the structural assignment obtained from XRD analysis.[20,21]

The morphology of the synthesized powders was subsequently examined by SEM (Figure 2e). The solid-state synthesis resulted in irregularly shaped micrometer-sized particles with an average size of approximately 2 μm. Elemental mapping performed by EDS (Figure 2f) revealed a homogeneous spatial distribution of K, Ga, Ge, O, Bi, and Eu within the analyzed particles, with no evident regions of elemental segregation. These observations, together with the XRD results, confirm the preparation of structurally homogeneous $KGaGeO_4$-based materials suitable for further investigation of their luminescence properties.

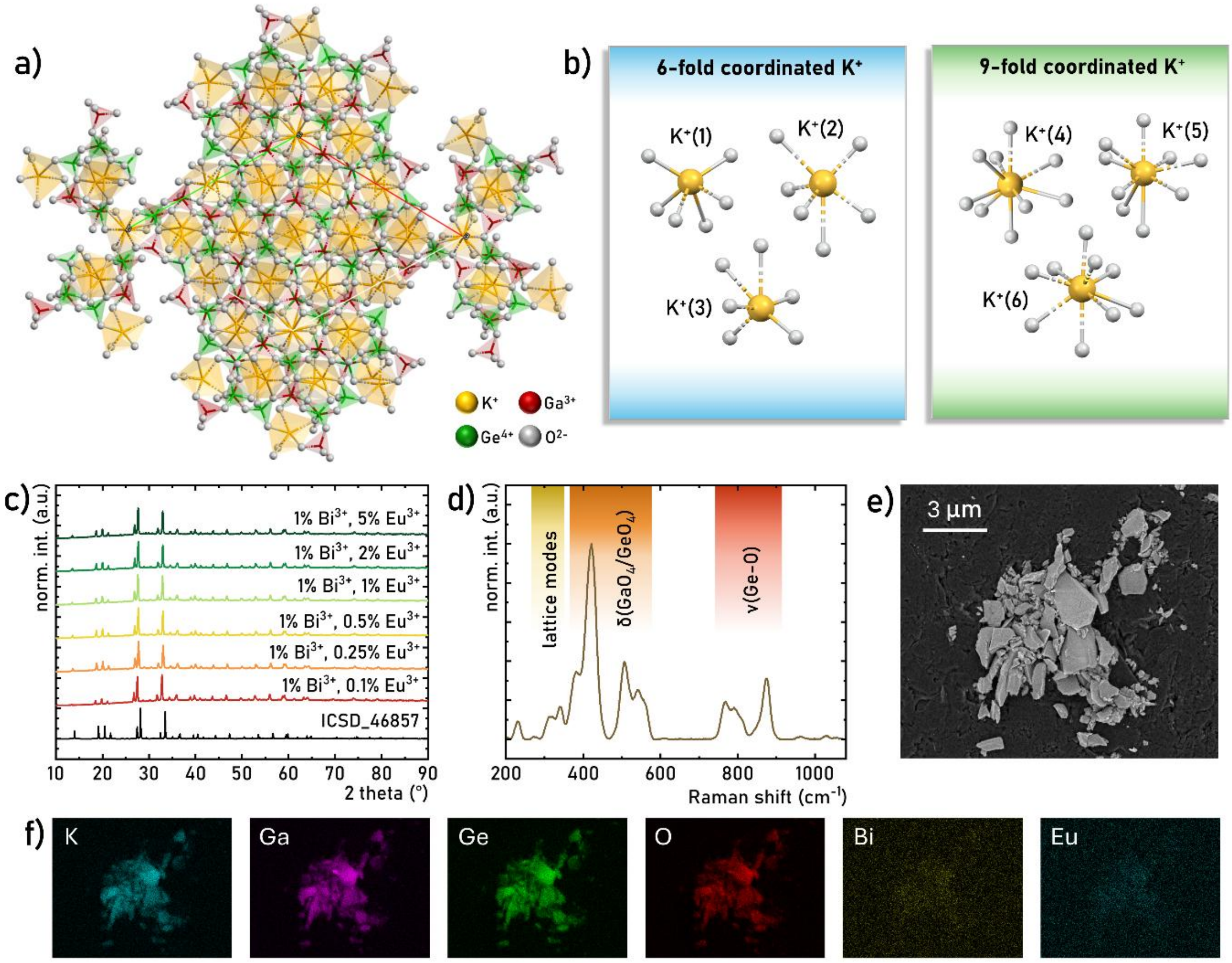

**Figure 2.** Crystal structure of $KGaGeO_4$ - a). Schematic representation of the two types of $K^+$ coordination polyhedra, with coordination numbers of 6 and 9 - b). XRD patterns of $KGaGeO_4$:1% $Bi^{3+}$, x%$Eu^{3+}$ (x = 0.1-5) - c). Raman spectrum of undoped $KGaGeO_4$ recorded at 93 K - d). SEM image of the $KGaGeO_4$:1%$Bi^{3+}$, 0.5%$Eu^{3+}$ - e), and the corresponding EDS elemental maps - f).

To gain deeper insight into the luminescence mechanisms operating in $KGaGeO_4$:$Bi^{3+}$, $Eu^{3+}$ the spectroscopic properties of the sample doped exclusively with 1% $Bi^{3+}$ were investigated first. The broad emission profile (recorded at 93 K under 300 nm excitation) can be satisfactorily decomposed by three Gaussian components centered at 23,573 $cm^{-1}$, 19,447 $cm^{-1}$ and 21,430 $cm^{-1}$ hereafter denoted as $Bi^{3+}$(1), $Bi^{3+}$(2) and matrix-related emission, respectively. The intermediate component was tentatively attributed to intrinsic emission of the $KGaGeO_4$ host, most likely associated with the defects.[22] To verify this assignment experimentally, the luminescence properties of $KGaGeO_4$ were additionally investigated, and the corresponding excitation and emission spectra are presented in the Supporting Information (Figure S1). The undoped host exhibits a broad emission band extending from approximately 400 to 650 nm, with a maximum centered at around 520 nm. Its spectral position and width overlap with the intermediate component obtained from the deconvolution of the $KGaGeO_4$:1%$Bi^{3+}$ emission spectrum, supporting its assignment to intrinsic matrix-related luminescence rather than to an additional $Bi^{3+}$ center. The corresponding excitation spectrum is also broad, exhibiting a maximum at around 270 nm. The presence of this intrinsic excitation pathway further indicates that the host lattice contributes directly to the overall luminescence response of $KGaGeO_4$:$Bi^{3+}$. Consequently, the experimentally observed emission spectrum should be considered as a superposition of three contributions originating from two spectroscopically distinguishable $Bi^{3+}$ centers and the intrinsic luminescence of the $KGaGeO_4$. The presence of two distinct emission components strongly suggests that $Bi^{3+}$ ions occupy at least two nonequivalent local environments within the $KGaGeO_4$ lattice.[22,23] Both bands exhibit pronounced spectral broadening, characteristic of $Bi^{3+}$ ions with an $ns^2$ electronic configuration, for which the emission is commonly associated with the $^3P_1 \rightarrow {}^1S_0$ transition.[24,25] The large bandwidth reflects the strong coupling of the excited $Bi^{3+}$ states with the surrounding lattice and the resulting distribution of local configurations. Further evidence for the presence of two spectroscopically distinct $Bi^{3+}$ centers is provided by the corresponding excitation spectra monitored at the maxima of the two emission components (Figure 3a). The excitation bands differ markedly in their spectral positions. For the higher-energy $Bi^{3+}$(1) emission, the excitation maximum is located at 33,624 $cm^{-1}$, whereas the excitation spectrum monitored for the lower-energy $Bi^{3+}$(2) emission reaches its maximum at 32,070 $cm^{-1}$. These broad excitation bands are assigned to the spin-orbit allowed $^1S_0 \rightarrow {}^3P_1$ transition of $Bi^{3+}$. Consequently, the estimated Stokes shifts amount to 10,051 $cm^{-1}$ and 10,640 $cm^{-1}$ for $Bi^{3+}$(1) and $Bi^{3+}$(2), respectively. The relatively small difference between these values suggests that both centers may experience local environments of comparable rigidity. Thus, the present spectroscopic results do not necessarily require the two $Bi^{3+}$ centers to occupy sites with different coordination numbers, but are also consistent with their incorporation into crystallographically nonequivalent $K^+$ positions characterized by the same coordination number. This interpretation can be considered in the context of the six crystallographically distinct $K^+$ sites present in the $KGaGeO_4$ structure. In particular, the three six-fold coordinated $K^+$(1-3) positions exhibit more pronounced differences in polyhedral distortion and local $K^+$-$O^{2-}$ geometry than the nine-fold coordinated $K^+$(4-6) sites. Consequently, the two $Bi^{3+}$ emission centers may plausibly originate from $Bi^{3+}$ ions incorporated into different six-coordinated $K^+$ sites, where subtle variations in local distortion, bond-length distribution, and coordination geometry could account for their distinct emission energies despite the similar Stokes shifts. However, assigning a particular $Bi^{3+}$ center to a specific $K^+$ position is not justified on the basis of the present data alone. Accordingly, the proposed site assignment should be regarded as a spectroscopically supported hypothesis rather than an unequivocal structural determination. Definitive identification of the crystallographic positions occupied by $Bi^{3+}$ would require complementary techniques sensitive to the local coordination environment. At the present stage, the combined structural and spectroscopic results indicate that the two $Bi^{3+}$ centers most likely originate from occupation of distinct $K^+$ environments, with the six-coordinated $K^+$(1-3) positions representing the most plausible candidates.

The coexistence of these two $Bi^{3+}$ centers has an important consequence for the optical properties of the material, since it enables excitation-wavelength-dependent tuning of the emission color. As shown in Figure 3b, excitation at wavelengths below approximately 300 nm preferentially enhances the higher-energy blue emission assigned to $Bi^{3+}$(1). At approximately 300 nm, the contributions of the two $Bi^{3+}$ centers become comparable, whereas excitation at longer wavelengths selectively favors the lower-energy green emission of $Bi^{3+}$(2). Thus, even in the absence of $Eu^{3+}$, the emission color of $KGaGeO_4$ can be continuously modified by changing the excitation wavelength.

The wavelength-dependent luminescence decay kinetics recorded over the 380-600 nm emission range (Figure 3d and Figure S2) reveal pronounced changes in the decay time across the broad emission profile, providing additional evidence for the coexistence of three spectroscopically distinguishable contributions. At the high-energy side of the emission band, the decay time reaches approximately 13 μs at 410 nm, where the contribution from $Bi^{3+}$(1) predominates. With increasing monitoring wavelength, the $\tau_{avr}$ decreases markedly, reaching a minimum of approximately 6.5 μs, when monitored at 460 nm. Notably, this wavelength closely corresponds to the maximum of the matrix-related component identified by spectral deconvolution at 21,430 $cm^{-1}$ (~467 nm), supporting its assignment to an emission channel characterized by distinctly faster decay kinetics. At longer wavelengths, the $\tau_{avr}$ increases again, reaching approximately 15 μs at 540 nm, where the contribution from the lower-energy $Bi^{3+}$(2) center becomes dominant. Thus, the non-monotonic evolution of the decay time across the emission spectrum is consistent with the superposition of three emitting components possessing different decay kinetics and independently supports the spectral deconvolution of the broad emission band into $Bi^{3+}$(1), host-related, and $Bi^{3+}$(2) contributions.

Having established the presence and spectroscopic behavior of two nonequivalent $Bi^{3+}$ centers, the effect of introducing $Eu^{3+}$ ions into the $KGaGeO_4$ system was subsequently investigated. For this purpose, $KGaGeO_4$:1% $Bi^{3+}$,x% $Eu^{3+}$ concentrations ranging from 0.1% to 5% were synthesized. Their emission spectra were recorded at 93 K under 340 nm excitation and are compared in Figure 3e. With increasing $Eu^{3+}$ concentration, the relative contribution of the $Bi^{3+}$ emission gradually decreases, whereas the characteristic narrow-line emissions of $Eu^{3+}$, associated with the $^5D_0 \rightarrow {}^7F_J$ (J = 0-4)[26] transitions become progressively more pronounced. Consequently, the overall emission color systematically shifts from the blue/green $Bi^{3+}$-dominated region toward the red spectral region characteristic of $Eu^{3+}$. This evolution is directly reflected in the corresponding CIE1931 chromaticity coordinates (Figure 3f), which change from $x$ = 0.319 and $y$ = 0.345 for the sample containing 0.1% $Eu^{3+}$ to $x$ = 0.551 and $y$ = 0.379 for the sample containing 5% $Eu^{3+}$.

The absolute photoluminescence quantum yield was also determined for all the co-doped samples (Figure 3g and Figure S3). At the 0.1% $Eu^{3+}$, the quantum yield reaches 13.6%. It subsequently increases to a maximum value of 45.9% for the $KGaGeO_4$:1% $Bi^{3+}$,0.25% $Eu^{3+}$. Further increase in the $Eu^{3+}$ concentration results in decrease in the quantum yield, reaching 10 % for the $KGaGeO_4$:1% $Bi^{3+}$,5% $Eu^{3+}$.

A schematic representation of the proposed luminescence mechanism in $KGaGeO_4$:$Bi^{3+}$,$Eu^{3+}$ is presented in Figure 3h. The experimental results indicate that both $Bi^{3+}$ centers can participate in the sensitization of $Eu^{3+}$, alongside excitation through the ligand-to-metal charge-transfer band, thereby providing additional pathways for populating the excited states of $Eu^{3+}$.[27,28] Evidence for this process is visible in the excitation spectra monitored at the $Eu^{3+}$ emission (Figure 3h). The characteristic $Eu^{3+}$ charge-transfer band, an additional broad contribution overlapping with the $Bi^{3+}$ A-bands assigned to the $^1S_0 \rightarrow {}^3P_1$ transition is clearly observed. The appearance of this $Bi^{3+}$-related excitation contribution while monitoring the $Eu^{3+}$ emission provides direct evidence of the $Bi^{3+} \rightarrow Eu^{3+}$ energy transfer. Additionally, the excitation spectrum monitored at the $Bi^{3+}$(2) emission contains a distinct high-energy band centered at approximately 280 nm, in addition to the main A-band. Since this feature is already present in the $KGaGeO_4$:$Bi^{3+}$ (Figure 3a), it cannot be attributed to the $Eu^{3+}$-related charge-transfer band or to a $Eu^{3+} \rightarrow Bi^{3+}$ back-energy-transfer process. Instead, its origin is more consistently associated with an intrinsic excitation pathway of the $KGaGeO_4$ host. This assignment is supported by the excitation spectrum of the $KGaGeO_4$ host (Figure S1), which exhibits a broad high-energy band in the same spectral region.

The contribution of the individual $Bi^{3+}$ centers to this process was further investigated for the sample containing 1% $Bi^{3+}$ and 0.1% $Eu^{3+}$ by recording emission spectra over a broad excitation range from 270 to 360 nm (Figure 3j). Within this excitation range, both $Bi^{3+}$ and $Eu^{3+}$ emission bands remain detectable; however, the relative contribution of the two $Bi^{3+}$ components varies strongly with excitation wavelength, following a trend analogous to that observed for the sample containing $Bi^{3+}$ alone. This confirms that the two $Bi^{3+}$ centers preserve their distinct excitation selectivity after $Eu^{3+}$ incorporation. Additional evidence for $Bi^{3+} \rightarrow Eu^{3+}$ energy transfer can be obtained by directly comparing the emission spectra of $KGaGeO_4:Bi^{3+}$ (Figure 3a and 3b) and $KGaGeO_4:Bi^{3+},Eu^{3+}$ recorded under identical 300 nm excitation. In the $Bi^{3+}/Eu^{3+}$-containing sample, the relative intensity of the lower-energy $Bi^{3+}(2)$ emission is reduced compared with that observed for $KGaGeO_4:Bi^{3+}$, suggesting depopulation of the corresponding $Bi^{3+}$ excited state through energy transfer to $Eu^{3+}$. This observation indicates that $Bi^{3+}(2)$ may act as a particularly effective sensitizing center for $Eu^{3+}$. The coexistence of two excitation-selective $Bi^{3+}$ centers together with narrow-line $Eu^{3+}$ emissions ultimately provides a wide range of excitation-dependent color tunability. As demonstrated by the CIE1931 diagram (Figure 3k), changing the excitation wavelength enables substantial displacement of the emission coordinates across the visible region. In particular, the emission can be tuned from a predominantly red output under 270 nm excitation to approximately white emission under 320 nm excitation. Therefore, the multichannel excitation and energy-transfer pathways present in $Bi^{3+}/Eu^{3+}$ co-doped $KGaGeO_4$ provide an effective strategy for excitation-controlled color modulation, arising from the combined contributions of two nonequivalent $Bi^{3+}$ centers and $Eu^{3+}$ emission.

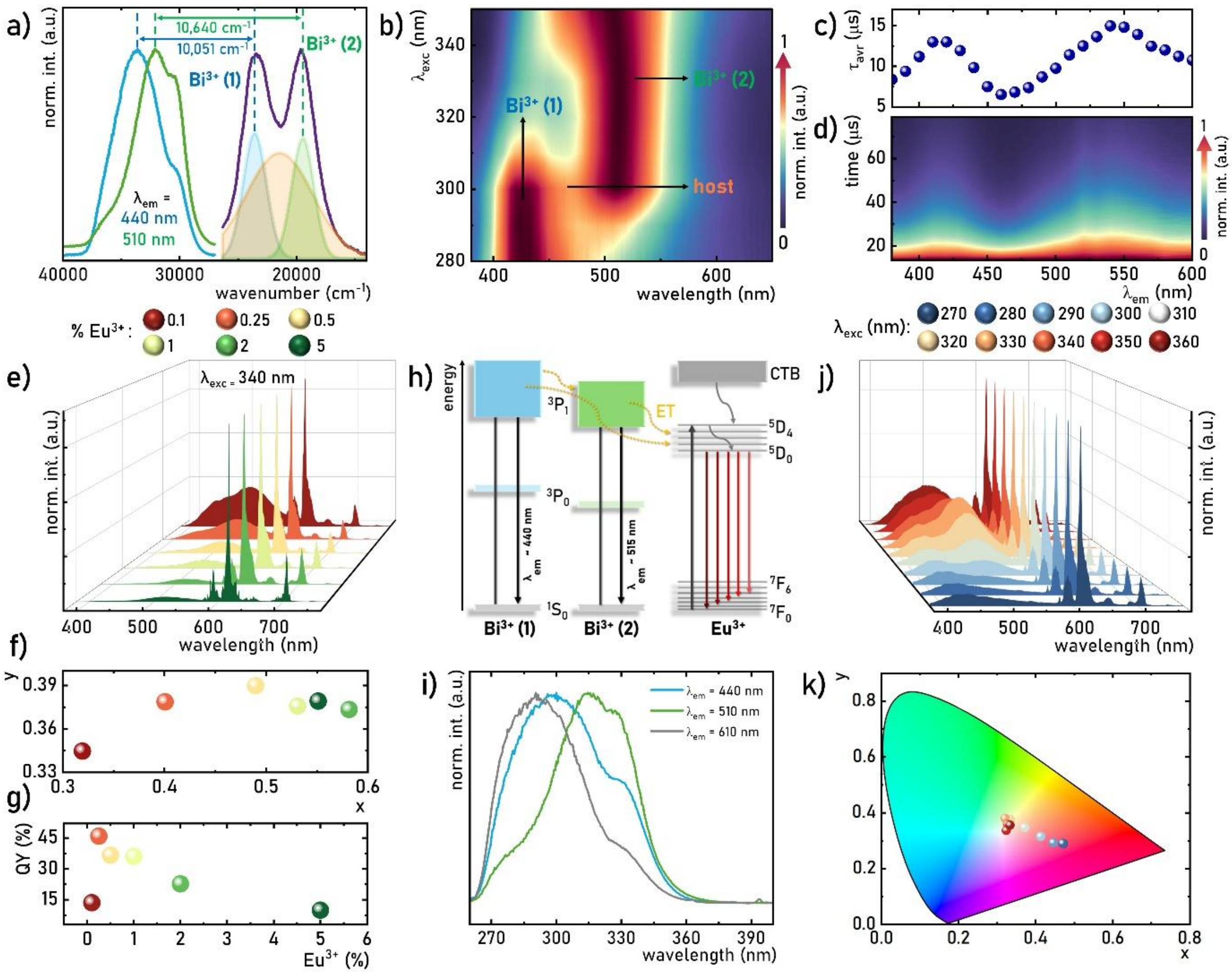

**Figure 3.** Emission spectrum recorded upon $\lambda_{exc}$ = 300 nm excitation together with excitation spectra monitored at the two $Bi^{3+}$-related emission maxima, $\lambda_{em}$ = 440 nm and 510 nm - a), emission spectra recorded as a function of excitation wavelength ($\lambda_{exc}$ = 280-350 nm) - b), $\tau_{avr}$ -c) and luminescence decay curves -d) as a function of monitored emission wavelength ($\lambda_{em}$ = 380-600 nm) measured at 93 K for $KGaGeO_4$:1%$Bi^{3+}$. Emission spectra recorded at 93 K - e), corresponding CIE1931 $x$ and $y$ chromaticity coordinates - f), and room-temperature quantum yield values - g) for $KGaGeO_4$:1%$Bi^{3+}$,x%$Eu^{3+}$ (x = 0.1-5). Schematic energy-level diagram of $Bi^{3+}$ and $Eu^{3+}$ illustrating the proposed energy-transfer pathways in $KGaGeO_4$:$Bi^{3+}$,$Eu^{3+}$- h). Excitation spectra monitored at selected emission maxima ($\lambda_{em}$ = 440 nm, 510 nm and 610 nm) - i), emission spectra recorded upon $\lambda_{exc}$ ranging from 270 to 360 nm at 93 K - j), and the corresponding CIE1931 chromaticity diagram recorded for for $KGaGeO_4$:1%$Bi^{3+}$, 0.1%$Eu^{3+}$ - k).

To identify the optimal excitation conditions for the development of a luminescent thermometer and to gain deeper insight into the excitation processes, a detailed analysis of the temperature-dependent spectroscopic properties of $KGaGeO_4$:1%$Bi^{3+}$,0.5%$Eu^{3+}$ was performed. Excitation spectra were recorded as a function of temperature while monitoring three representative emission wavelengths, namely 550, 440, and 611 nm (Figures 4a-c), enabling the observation of the evolution of the excitation channels associated with the individual luminescent centers to be followed. When monitoring the emission at 550 nm, a pronounced temperature-induced enhancement of the excitation intensity is observed in the approximately 320-360 nm region (Figure 4a). This spectral range is predominantly associated with the A-band excitation of the $Bi^{3+}$(2) center. A similar temperature evolution is observed for the excitation spectra monitored at 440 nm (Figure 4b). Although this wavelength is primarily associated with the $Bi^{3+}$(1) emission, the broad emission bands of the two $Bi^{3+}$ centers partially overlap. Consequently, the increasing contribution of the $Bi^{3+}$(2) center at elevated temperatures is also reflected in the excitation spectra recorded at 440 nm. In contrast, the excitation spectra monitored at the characteristic $Eu^{3+}$ emission at 611 nm exhibit a progressive decrease in intensity with increasing temperature (Figure 4c), indicating a gradual reduction in the efficiency of the excitation pathways feeding the $Eu^{3+}$ emitting state. To obtain further insight into the origin of these contrasting temperature dependences, the excitation spectra monitored at 550 nm were deconvoluted into three components, assigned to the A-band of $Bi^{3+}$(1), the A-band of $Bi^{3+}$(2), and a higher-energy host-related contribution (Figure 4d). The individual components exhibit markedly different thermal evolution. The $Bi^{3+}$(1)-related excitation band decreases monotonically with increasing temperature, reaching approximately 50% of its initial intensity at around 533 K. The host-related contribution also decreases progressively and becomes negligible at 233 K. In marked contrast, the $Bi^{3+}$(2)-related A-band exhibits a strong non-monotonic temperature dependence. Its intensity initially increases significantly, reaching a maximum of approximately 4.4 times its value at 93 K at around 493 K, and subsequently decreases at higher temperatures. The opposite thermal evolution of the $Bi^{3+}$(1) and $Bi^{3+}$(2) excitation channels points to a temperature-induced redistribution of excitation efficiency between the two $Bi^{3+}$ centers. In particular, the pronounced enhancement of the $Bi^{3+}$(2)-related excitation band suggests that population of this center becomes increasingly favorable upon heating. This behavior is consistent with the simultaneous increase in the corresponding broad $Bi^{3+}$(2) emission observed over a considerable part of the investigated temperature range (Figure 4e). Therefore, the unusual increase in $Bi^{3+}$(2) emission intensity should not be considered solely in terms of conventional thermal quenching, but rather as a consequence of a thermally modified excitation pathway and redistribution of excitation energy among the available luminescent centers. Because the excitation channels associated with $Bi^{3+}$ and $Eu^{3+}$ exhibit markedly different temperature dependences, the resulting emission response is expected to strongly depend on the excitation wavelength. To investigate this effect systematically, temperature-dependent emission spectra were therefore recorded upon excitation at 300, 320, and 340 nm, and the corresponding spectral maps are presented in Figure 4e. Pronounced differences in the relative contributions of the broad $Bi^{3+}$ emission bands and the characteristic $Eu^{3+}$ emission lines are observed for the three excitation wavelengths. As a result, both the spectral distribution and the emission color vary substantially with temperature and excitation wavelength. This behavior is clearly reflected in the CIE1931 chromaticity diagram shown in Figure 4f, where distinct temperature-dependent color trajectories are observed for the three excitation wavelengths. The emission evolves over a broad range of chromaticity coordinates, from red/orange to turquoise, depending on the selected excitation conditions and temperature. These changes are also directly visible in the luminescence photographs recorded at 93 and 673 K upon excitation at 300, 320, and 340 nm (Figure 4g). To quantitatively describe the thermally induced redistribution of emission intensity between the $Bi^{3+}$- and $Eu^{3+}$-dominated spectral regions, the luminescence intensity ratio, $LIR_1$, was calculated according to the following equation:

$$LIR_1 = \frac{\int_{456nm}^{570nm} I(Bi^{3+}(1,2): {}^3P_1 \rightarrow {}^1S_0) d\lambda}{\int_{575nm}^{720nm} I(Eu^{3+}: {}^5D_0 \rightarrow {}^7F_J) d\lambda} \quad \textbf{(4)}$$

At 93 K, the $LIR_1$ values were 0.30, 0.69, and 0.49 for excitation wavelengths of 300, 320, and 340 nm, respectively (Figure 4h). Upon increasing the temperature to 633 K, these values increased approximately 8-, 6-, and 11-fold, respectively. Thus, the largest relative modulation of the $LIR_1$ was obtained under excitation at 340 nm. The thermometric performance was subsequently evaluated using the relative thermal sensitivity, $S_{R(T)}$, defined as:

$$S_{R(T)} = \frac{1}{X}\frac{\Delta X}{\Delta T}100\% \quad \textbf{(5)}$$

where $X$ represents the temperature-dependent luminescence parameter and d $\Delta X / \Delta T$ describes its variation with temperature. The highest relative sensitivity was obtained for excitation at 340 nm, reaching 0.78 % $K^{-1}$ at 143 K. For excitation at 300 and 320 nm, the maximum $S_{R(T)}$ values were 0.56 and 0.47 % $K^{-1}$, observed at 411 and 381 K, respectively (Figure 4i). Considering the highest maximum relative sensitivity, excitation at 340 nm was selected as the most suitable excitation condition for further temperature-dependent investigations.

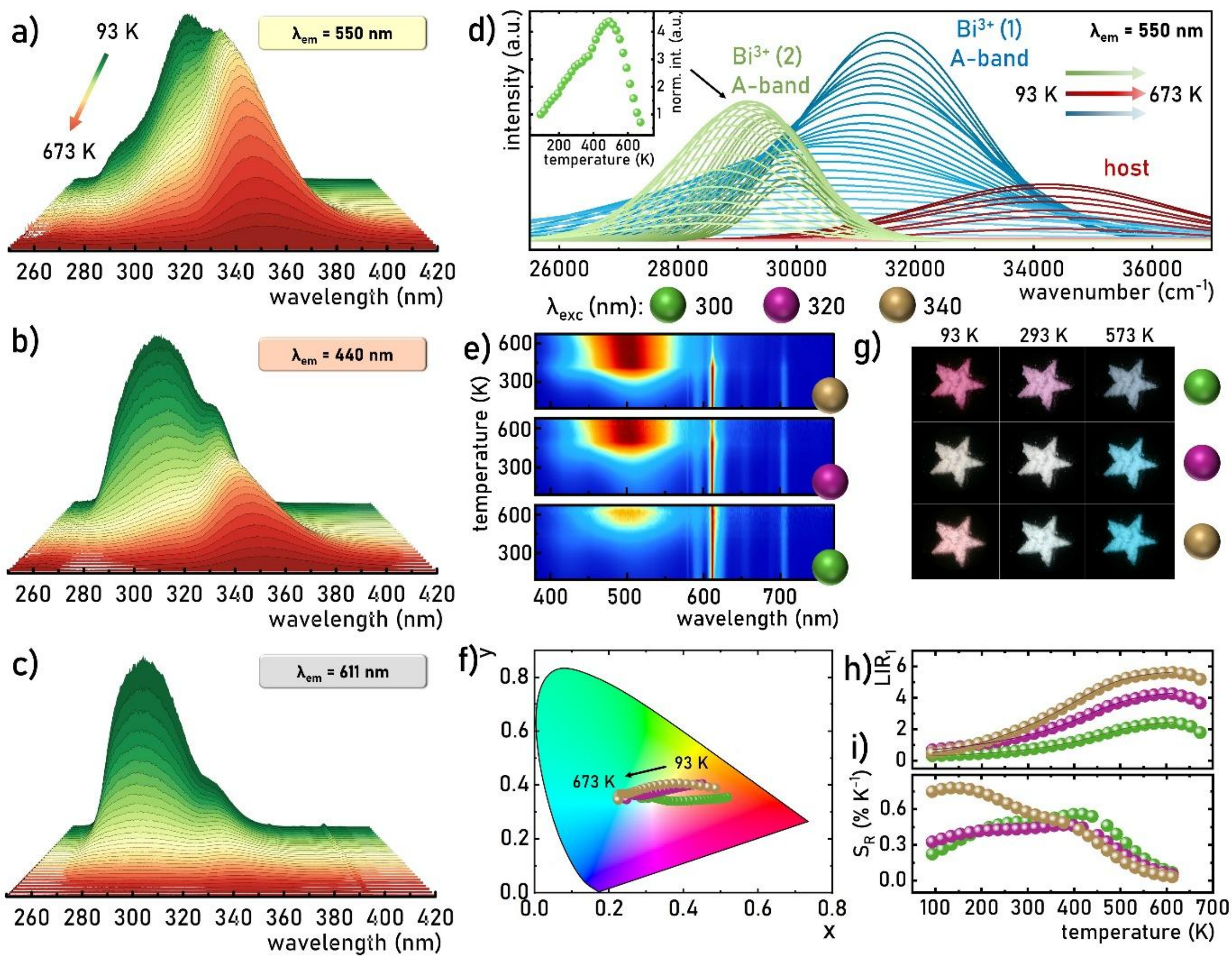

**Figure 4.** Temperature-dependent excitation spectra monitored at $\lambda_{em}$ = 550 nm - a), 440 nm - b), and 611 nm - c). Temperature-dependent deconvolution of the excitation spectra monitored at $\lambda_{em}$ = 550 nm - d). Temperature-dependent emission spectra recorded upon different excitation wavelengths, $\lambda_{exc}$ = 300, 320, and 340 nm - e), CIE1931 chromaticity coordinates as a function of temperature - f). Photographs of the luminescence of $KGaGeO_4$:1%$Bi^{3+}$,0.5%$Eu^{3+}$ recorded at 93, 293 and 573 K under different excitation wavelengths - g). Thermal dependance of $LIR_1$ - h) and corresponding $S_R$ - i), and for $KGaGeO_4$:1%$Bi^{3+}$,0.5%$Eu^{3+}$ under different excitation wavelengths.

All samples were investigated as a function of temperature to evaluate their potential for luminescence thermometry. Figures 5a and 5b present representative emission spectra recorded as a function of temperature for the two limiting compositions, $KGaGeO_4$:1%$Bi^{3+}$,0.1%$Eu^{3+}$ and $KGaGeO_4$:1%$Bi^{3+}$,5%$Eu^{3+}$, while the corresponding spectra for the intermediate $Eu^{3+}$ concentrations are provided in Figure S4. Both compositions exhibit a pronounced temperature-induced redistribution of emission intensity between the broad $Bi^{3+}$-related bands and the characteristic $Eu^{3+}$ emission lines. At low temperature, the contribution of $Eu^{3+}$ emission is particularly strong, especially for the highly $Eu^{3+}$-doped sample. Upon heating, however, the $Eu^{3+}$ emission gradually weakens, whereas the contribution of the lower-energy $Bi^{3+}$(2) band increases markedly. This contrasting thermal response is also clearly reflected in the two-dimensional temperature-dependent emission maps shown in the insets of Figures 5a and 5b. The observed behavior points to a competition between thermally activated population of the $Bi^{3+}$(2)-related excited state and thermal depopulation or quenching of the $Eu^{3+}$ $^5D_0$ emitting level. In particular, the enhancement of the $Bi^{3+}$(2) emission is consistent with the temperature-induced increase in the corresponding excitation-band intensity shown in Figures 4a and 4d, suggesting that heating facilitates population of this lower-energy $Bi^{3+}$-related excited state. This may arise from thermally assisted redistribution between the two $Bi^{3+}$ centers and/or from activation of additional excitation pathways feeding the $Bi^{3+}$(2) state. At the same time, the $Eu^{3+}$ emission becomes progressively less competitive because of increasing non-radiative losses and a reduced relative contribution of the $Eu^{3+}$ emission channel. To quantify this spectral redistribution, the broad $Bi^{3+}$ emission was deconvoluted into contributions originating from two spectroscopically distinct $Bi^{3+}$ centers, denoted $Bi^{3+}$(1) and $Bi^{3+}$(2). The resulting relative spectral contributions provide a quantitative measure of how the emission balance evolves with temperature. For $KGaGeO_4$:1%$Bi^{3+}$,0.1%$Eu^{3+}$ at 93 K, the relative contributions of $Bi^{3+}$(1), $Bi^{3+}$(2), and $Eu^{3+}$ amount to approximately 12.3%, 62.6%, and 25.1%, respectively (Figure 5c). In contrast, for $KGaGeO_4$:1%$Bi^{3+}$,5%$Eu^{3+}$, the corresponding values are approximately 1.0%, 20.1%, and 78.9%. Importantly, increasing temperature drives both systems toward a $Bi^{3+}$(2)-dominated emission regime, although the trajectory of this redistribution is substantially more pronounced for the highly $Eu^{3+}$-doped sample. The ternary representation of the relative $Bi^{3+}$(1), $Bi^{3+}$(2), and $Eu^{3+}$ emission contributions further emphasizes that temperature does not simply reduce the overall emission intensity, but continuously redistributes the relative contribution of the spectroscopically distinct emissive channels. The opposite temperature dependences of the $Bi^{3+}$(2) and $Eu^{3+}$ emission channels result in a substantial change in the chromaticity coordinates (Figure 5d). For $KGaGeO_4$:1%$Bi^{3+}$,0.1%$Eu^{3+}$, heating from 93 to 673 K produces a moderate color evolution from nearly white emission toward light blue. In contrast, $KGaGeO_4$:1%$Bi^{3+}$,5%$Eu^{3+}$ exhibits a much more pronounced color transformation, from red-dominated emission at low temperature to light-blue emission at high temperature. For the 0.1% $Eu^{3+}$ co-doped sample, the $x$ coordinate changes monotonically from 0.32 to 0.26 over the 93-673 K range, whereas for the 5% $Eu^{3+}$ co-doped sample a significantly larger decrease is observed, from 0.54 to 0.25, corresponding to a relative variation of approximately 50% (Figure 5e). The $y$ coordinate, however, exhibits non-monotonic behavior for all compositions (Figure S5): it initially increases upon heating, subsequently decreases in the intermediate-temperature range, and rises again at higher temperatures. Such behavior most likely reflects the simultaneous and non-equivalent evolution of several overlapping emission contributions and therefore prevents the $y$ coordinate from serving as a robust single-valued thermometric parameter. Consequently, only the $x$ coordinate was used to evaluate the chromaticity-based thermometric performance. The relative sensitivity, calculated according to Eq. 5, reaches its highest value for the 5% $Eu^{3+}$ co-doped sample, with a maximum of 0.32% $K^{-1}$ at 308 K (Figure 5f). Lower but still significant sensitivities are observed for the remaining compositions, for example 0.19% $K^{-1}$ at 93 K for the 0.1% $Eu^{3+}$ sample and 0.15% $K^{-1}$ at 437 K for the 2% $Eu^{3+}$ sample. These results demonstrate that CIE1931-coordinate thermometry can provide an additional optical readout channel. However, because chromaticity coordinates are derived from the entire emission spectrum and may be influenced by spectral overlap, detector response, and color-space transformation, a ratiometric intensity-based approach was further explored as a more direct and potentially more robust thermometric strategy. To fully exploit the thermometric potential of $KGaGeO_4:Bi^{3+},Eu^{3+}$, the temperature dependence of the luminescence intensity ratio was therefore investigated. Based on the temperature-induced spectral changes, two narrow spectral windows exhibiting strongly contrasting thermal behavior were selected: the $Bi^{3+}$-related region at 459-464 nm and the $Eu^{3+}$ $^5D_0 \rightarrow ^7F_2$ emission region at 615.5-620.5 nm. Their temperature dependences are shown in Figures 5g and 5h, respectively. The intensity of the $Bi^{3+}$-related spectral window initially increases with temperature, and the magnitude of this enhancement depends strongly on $Eu^{3+}$ concentration. For $Eu^{3+}$ concentrations up to 3%, the thermally induced enhancement becomes progressively more pronounced, reaching its largest value for the 2% $Eu^{3+}$ co-doped sample, for which the intensity increases by approximately 6.5 times between 93 and 453 K. For the 5% $Eu^{3+}$ composition, the enhancement is weaker, reaching approximately 3.7 times at 433 K. This behavior suggests that the efficiency of thermally assisted population of the $Bi^{3+}$-related emitting state is strongly influenced by $Eu^{3+}$ concentration and by the competition between $Bi^{3+}$-centered emission and $Bi^{3+} \rightarrow Eu^{3+}$ energy-transfer pathways. In contrast, the intensity of the selected $Eu^{3+}$ emission window decreases monotonically with temperature for most compositions. The only exception is the 5% $Eu^{3+}$ co-doped sample, for which a small initial increase of approximately 12% is observed up to 173 K before thermal quenching becomes dominant. This low-temperature enhancement may indicate a redistribution of excitation energy toward $Eu^{3+}$ at the initial stage of heating, before non-radiative losses become dominant at higher temperatures. The strongly opposite thermal responses of the selected $Bi^{3+}$ and $Eu^{3+}$ spectral regions make their intensity ratio particularly suitable for ratiometric thermometry. Accordingly, $LIR_2$ was calculated as:

$$LIR_2 = \frac{\int_{459nm}^{464nm} I(Bi^{3+}: {}^3P_1 \rightarrow {}^1S_0)d\lambda}{\int_{615.5nm}^{620.5nm} I(Eu^{3+}: {}^5D_0 \rightarrow {}^7F_2)d\lambda} \quad \textbf{(6)}$$

The $LIR_2$ values at 93 K decrease systematically with increasing $Eu^{3+}$ concentration, directly reflecting the changing balance between the $Bi^{3+}$- and $Eu^{3+}$-related emission contributions (Figure 5i). Upon heating, the $LIR_2$ increases for all compositions as a consequence of the opposite thermal evolution of the two selected spectral regions: the $Bi^{3+}$-related intensity increases, whereas the $Eu^{3+}$ $^5D_0 \rightarrow ^7F_2$ emission progressively decreases. The increase in $LIR_2$ is maintained up to approximately 600 K for most compositions. For $KGaGeO_4$:1%$Bi^{3+}$,0.1%$Eu^{3+}$, the $LIR_2$ increases from 0.71 at 93 K to a maximum of 4.29 at 593 K, whereas for $KGaGeO_4$:1%$Bi^{3+}$,5%$Eu^{3+}$ it rises from only 0.023 at 93 K to 1.25 at 613 K. Although the absolute $LIR_2$ values are lower for the highly $Eu^{3+}$-doped sample, its relative variation with temperature is considerably stronger. This is consistent with the pronounced temperature-driven redistribution of the emission balance observed for this composition, where the low-temperature spectrum is strongly dominated by $Eu^{3+}$ emission, while heating progressively shifts the emission toward the $Bi^{3+}$(2)-related channel. Because the $LIR_2$ response remains monotonic over a broad temperature interval, the working range of the ratiometric thermometer was defined as 93-613 K. The relative sensitivity was calculated according to Eq. 5. Among all investigated compositions, the $KGaGeO_4$:1%$Bi^{3+}$,5%$Eu^{3+}$ phosphor exhibits the best thermometric performance, reaching a maximum relative sensitivity of 1.55% $K^{-1}$ at 299 K (Figures 5j and S6). Importantly, $S_{R(T)}$ remains above 1% $K^{-1}$ over a broad temperature interval from 213 to 393 K. This behavior can be attributed to the particularly strong contrast between the thermally enhanced $Bi^{3+}$-related emission and the simultaneously quenched $Eu^{3+}$ emission in this composition. The $KGaGeO_4$:1%$Bi^{3+}$,5%$Eu^{3+}$ provides the most favorable combination of a strongly $Eu^{3+}$-dominated low-temperature spectrum and a pronounced thermally induced shift toward $Bi^{3+}$ emission, resulting in the largest relative change in $LIR_2$ and, consequently, the highest $S_{R(T)}$. Taking into account both its superior ratiometric sensitivity and the largest temperature-induced chromaticity variation among the investigated compositions, $KGaGeO_4$:1%$Bi^{3+}$,5%$Eu^{3+}$ was selected for further demonstration of its temperature-sensing capability.

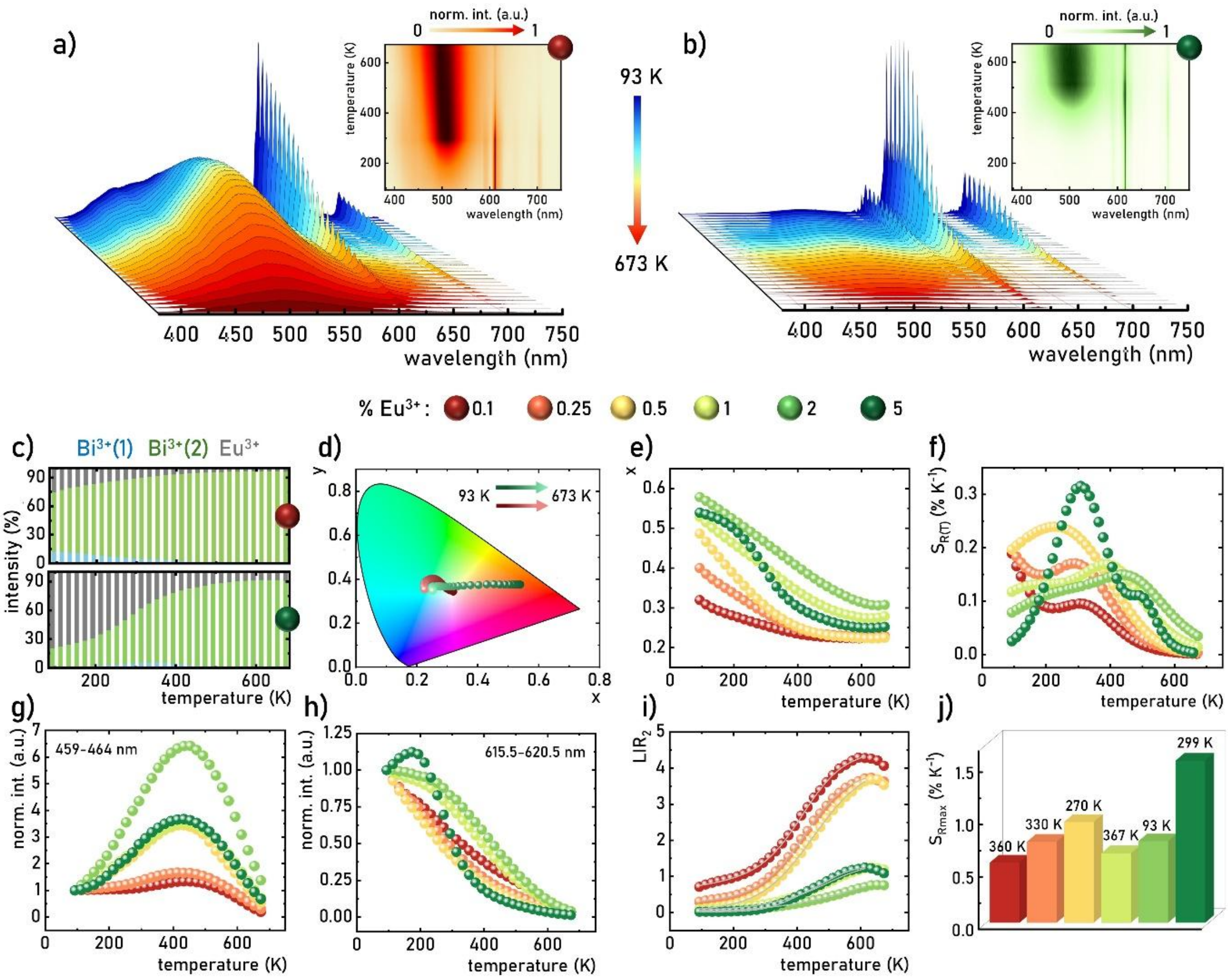

**Figure 5.** Temperature-dependent emission spectra ($\lambda_{exc}$ = 340 nm) and corresponding normalized emission intensity maps for $KGaGeO_4$:1%$Bi^{3+}$,0.1%$Eu^{3+}$ - a) and $KGaGeO_4$:1%$Bi^{3+}$,5%$Eu^{3+}$ - b). Thermal evolution of the relative contributions of $Bi^{3+}$(1), $Bi^{3+}$(2), and $Eu^{3+}$ emission components in the emission spectra - c), and CIE1931 chromaticity coordinate x - d), for $KGaGeO_4$:1%$Bi^{3+}$,0.1%$Eu^{3+}$ and $KGaGeO_4$:1%$Bi^{3+}$,5%$Eu^{3+}$. Thermal evolution of the CIE1931 x-coordinate - e), corresponding $S_{Rx}$ - f), normalized emission intensity integrated over the 459-464 nm range - g) and 615.5-620.5 nm range - h), and *LIR* - i), for $KGaGeO_4$:1%$Bi^{3+}$,x$Eu^{3+}$. Comparison of the maximum $S_R$ and the corresponding temperatures for all investigated samples - j).

Luminescence photographs of $KGaGeO_4$:1%$Bi^{3+}$,5%$Eu^{3+}$ were recorded under 340 nm excitation over the 93-673 K range to evaluate the suitability of the material for RGB-based temperature readout and spatial temperature imaging (Figures 6 and S7). The photographs reveal a pronounced evolution of the emission color from red through white to blue with increasing temperature. This visual change directly reflects the temperature-induced redistribution of the emission intensity between the $Eu^{3+}$ and $Bi^{3+}$ components, as evidenced by the progressive increase in the $Bi^{3+}/Eu^{3+}$ intensity ratio in the corresponding emission spectra (Figure 6b). To quantify the observed color evolution, the photographs were decomposed into the red (R), green (G), and blue (B) channels (Figure S8). The individual channels exhibit markedly different temperature dependences. The normalized G and B intensities increase with temperature up to approximately 413 K, reaching about 1.6 and more than 2 times their values at 93 K, respectively. In contrast, the R-channel intensity decreases strongly with increasing temperature and is reduced by more than one order of magnitude at 673 K relative to its initial value (Figures 6c and S8). These contrasting trends provide a favorable basis for ratiometric analysis, and the R/G and R/B channel ratios were therefore selected as thermometric parameters (Figures 6d and 6e). Both ratios decrease with increasing temperature and yield $S_{R(T)}$, of 0.75 and 1.05 % $K^{-1}$, respectively. Within the temperature interval selected for further analysis namely 293-633 K, both parameters exhibit an approximately linear response, allowing them to be implemented simultaneously in a multiple linear regression (MLR) model. MLR has previously been applied in luminescence thermometry and manometry to combine several temperature/pressure-dependent observables within a single calibration model and thereby improve sensing performance.[10,29–32] In the present case, R/G and R/B were used as predictors, whereas temperature was treated as the response variable:

$$p = \beta_0 + \beta_1 \left(\frac{R}{G}\right) + \beta_2 \left(\frac{R}{B}\right) + \varepsilon \qquad \textbf{(7)}$$

where $\beta_0$ is the intercept, $\beta_1$ and $\beta_2$ are the regression coefficients associated with the ratiometric parameters, and $\varepsilon$ is the residual error. The model was constructed following previously reported procedures.[10,32] The relative importance of the two predictors was strongly asymmetric, with R/G carrying most of the information used by the model (~90%), while R/B provided a smaller complementary contribution (~10%) (inset of Figure 6f and Table S1). The simultaneous use of both mentioned RGB ratios substantially improved the thermometric performance. The maximum $S_{R(T)}$ increased to 8.82% $K^{-1}$, corresponding to an almost ninefold enhancement relative to the R/B ratio considered individually (Figure 6f). This result demonstrates that multivariate treatment can extract additional temperature-dependent information from simple luminescence photographs beyond that available from a single RGB ratio. The excellent agreement between the experimentally applied and MLR-predicted temperatures further confirms the potential of this approach for quantitative RGB-based thermometry and, ultimately, spatially resolved temperature imaging (Figure 6g). An important practical advantage of this methodology is that it does not require sophisticated spectroscopic instrumentation, since the relevant optical information can be extracted directly from conventional RGB images. The combination of straightforward image acquisition with computational post-processing therefore provides an accessible and readily implementable sensing strategy, while multivariate analysis substantially improves the sensing performance.

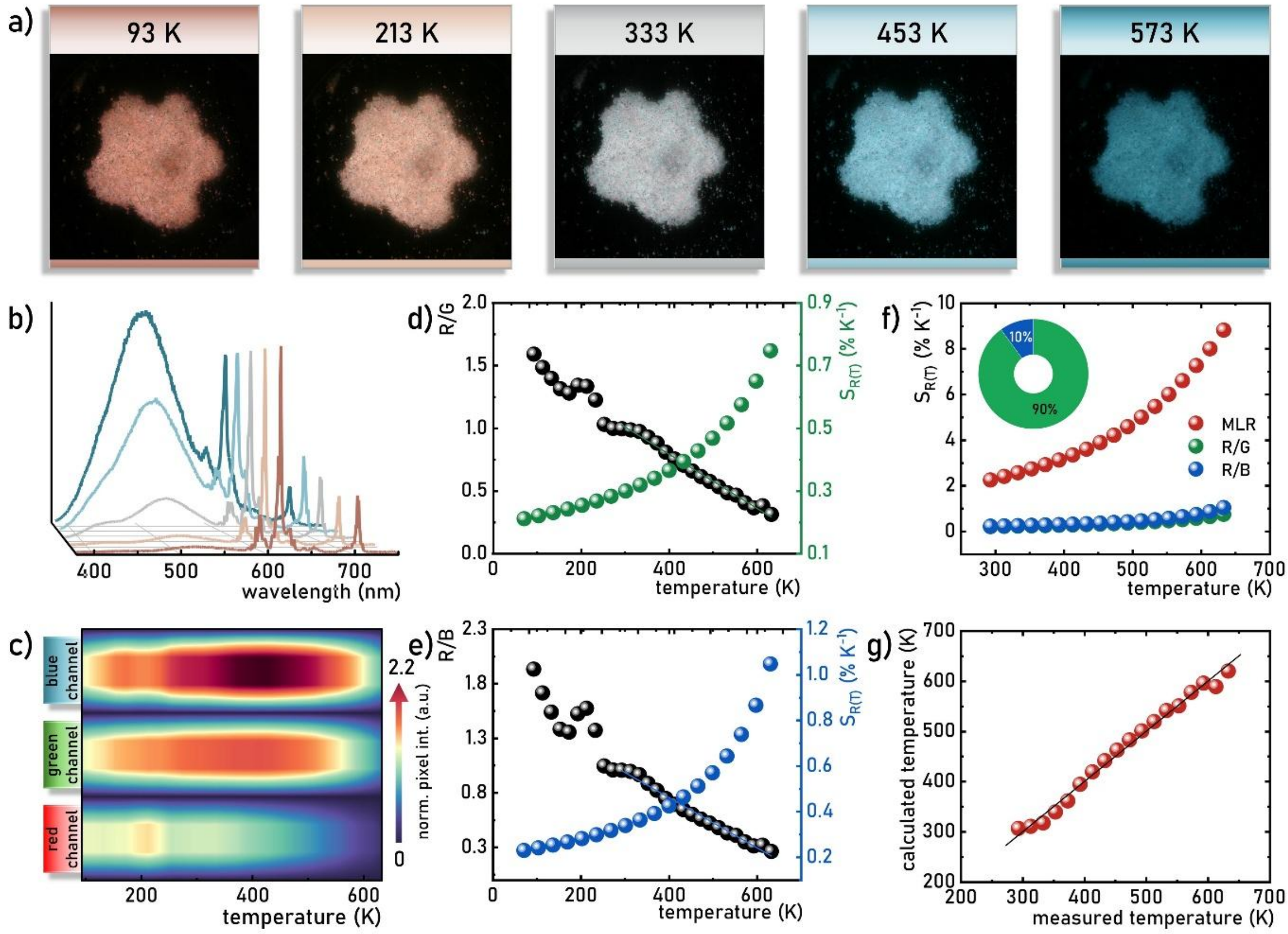

**Figure 6.** Luminescence photographs - a) and emission spectra -b) obtained upon 340 nm excitation at selected temperatures for $KGaGeO_4$:1%$Bi^{3+}$,5%$Eu^{3+}$. Temperature evolution of the normalised red (R), green (G), and blue (B) color-channel intensities extracted from the luminescence images - c). Temperature dependence of the R/G - d) and R/B - e) channel-intensity ratios together with the corresponding $S_{R(T)}$. Comparison of the $S_{R(T)}$ obtained from the RGB-based ratiometric parameters and from the MLR model with inset of the values correspond to the standardized $\beta$-weights - f). Comparison between the temperature predicted by the MLR model and the experimentally applied temperature - g).

Based on the pronounced and composition-dependent thermal response of $KGaGeO_4$:$Bi^{3+}$,$Eu^{3+}$, its luminescence behavior under compression was subsequently investigated to evaluate the potential of this material for optical pressure sensing. Since the $KGaGeO_4$:$Bi^{3+}$,0.1%$Eu^{3+}$ composition exhibited only a relatively weak temperature-induced color shift and comparatively low thermometric sensitivity, this sample was selected as the most promising candidate for manometric studies, where minimization of temperature cross-sensitivity is particularly desirable. Pressure-dependent measurements were performed over the 0-16 GPa range upon $\lambda_{exc}$ = 355 nm. The corresponding emission spectra, normalized to the $Eu^{3+}$ emission intensity, are presented in Figure 7a, while the two-dimensional pressure-dependent emission map is shown in Figure 7b. Under ambient conditions, the spectrum is dominated by the $Bi^{3+}$-related emission band, resulting in intense turquoise luminescence, as illustrated by the photographs recorded inside the DAC. Upon compression, the $Bi^{3+}$ emission decreases considerably faster than the $Eu^{3+}$ emission, progressively shifting the spectral balance toward the $Eu^{3+}$-related contribution. This contrasting pressure response can be rationalized by the different electronic structures of the two activators. The excited states of $Bi^{3+}$, originating from the $6s^2$ configuration and involving $6s6p$ states, are strongly influenced by changes in the local coordination environment, crystal-field strength, covalency, and lattice distortion induced by compression. In contrast, the $4f$ electrons of $Eu^{3+}$ are effectively shielded by the outer $5s$ and $5p$ shells, making the intra-configurational $4f$-$4f$ electronic transitions substantially less sensitive to changes in the surrounding crystal field.[26,33] Consequently, increasing pressure produces a much stronger perturbation of the $Bi^{3+}$-related emission than of the $Eu^{3+}$ emission, which can therefore serve as a comparatively pressure-stable internal spectral reference.[34,35] As a consequence of this progressive redistribution of emission intensity, the observed luminescence color changes markedly with pressure, evolving from turquoise under ambient conditions, through nearly white emission at approximately 5.2 GPa, to red emission at 14.7 GPa. This evolution is also reflected in the CIE1931 chromaticity diagram shown in Figure 7c, where the coordinates change from approximately $x$ = 0.25 and $y$ = 0.38 at ambient pressure to $x$ = 0.46 and $y$ = 0.36 at 16 GPa. In addition to the pronounced change in the relative $Bi^{3+}/Eu^{3+}$ emission intensities, compression also affects the spectral characteristics of the $Eu^{3+}$ emission. The position of the main $Eu^{3+}$ emission maximum shifts from 610.9 nm at ambient pressure to 614.7 nm at 16 GPa, while its FWHM increases from 9.7 to 17.5 nm. Although the 4*f*-4*f* transitions of $Eu^{3+}$ are relatively insensitive to the crystal field, pressure-induced changes in local symmetry and Stark splitting can still lead to measurable spectral shifts and line broadening. The observed increase in FWHM may therefore reflect enhanced structural distortion and a broader distribution of slightly different $Eu^{3+}$ local environments under compression.[36,37] Owing to the pronounced pressure-induced reshaping of the entire emission profile, the spectral centroid was selected as the first quantitative manometric parameter. As shown in Figure 7d, the centroid shifts monotonically from 514.4 nm at ambient pressure to 574.5 nm at 14.7 GPa, followed by a tendency toward saturation at higher pressures. Therefore, the useful pressure range for centroid-based sensing was defined as 0-14.7 GPa. The absolute sensitivity, $S_A$, was calculated according to:

$$S_A = \frac{\Delta X}{\Delta p} \qquad \textbf{(8)}$$

where $\Delta X$ represents the change of pressure-dependent spectroscopic parameter, in this case the spectral centroid, and $\Delta p$ is pressure change. A high $S_A$ exceeding 4 nm $GPa^{-1}$ was maintained over the entire useful pressure range, with a maximum value of 4.56 nm $GPa^{-1}$ at 5.32 GPa. The pressure-induced color evolution provides a second, independent manometric readout based on the CIE1931 chromaticity coordinates. The *x*-coordinate increases monotonically with pressure, as shown in Figure 7e, and was therefore selected as the calibration parameter, whereas the *y*-coordinate exhibits a less regular pressure dependence (Figure S9). The relative sensitivity was calculated according to:

$$S_{R(p)} = \frac{1}{X}\frac{\Delta X}{\Delta p}100\% \qquad \textbf{(9)}$$

where $\Delta X$ denotes the selected pressure-dependent optical parameter and its change in pressure $\Delta p$. For the CIE1931 *x*-coordinate, the relative pressure sensitivity reaches 9.93% $GPa^{-1}$ under ambient conditions and gradually decreases to 1.77% $GPa^{-1}$ at 13.2 GPa. Thus, chromaticity-based sensing provides particularly high pressure sensitivity in the low-pressure region and additionally enables direct visual assessment of the applied pressure through the emitted color. Finally, the strongest manometric response was obtained using the ratiometric mode based on the relative $Bi^{3+}$ and $Eu^{3+}$ emission intensities. The $LIR_3$ was calculated according to Eq. 10 as:

$$LIR_3 = \frac{\int_{625nm}^{630nm} I(Eu^{3+} : {}^5D_0 \rightarrow {}^7F_2) d\lambda}{\int_{500nm}^{505nm} I(Bi^{3+} : {}^3P_1 \rightarrow {}^1S_0) d\lambda} \qquad \textbf{(10)}$$

The $LIR_3$ increases monotonically from 0.13 at ambient pressure to 2.94 at 14.7 GPa, after which the response approaches saturation (Figure 6f). Accordingly, the useful pressure range of the $LIR_3$-based manometric readout was defined as 0-14.7 GPa. The corresponding relative sensitivity, calculated using Eq. 8, is also presented in Figure 7f. A remarkably high maximum $S_{R(p)}$ of 54.47% $GPa^{-1}$ is obtained under ambient conditions. Although the sensitivity gradually decreases with increasing pressure, it remains high over the entire useful range and still reaches 4.65% $GPa^{-1}$ at 14.7 GPa. Such a large sensitivity originates from the strongly differential pressure response of the two emission channels: $Bi^{3+}$ is rapidly quenched upon compression, whereas $Eu^{3+}$ remains comparatively stable. Consequently, even a small pressure increase produces a pronounced relative change in the intensity ratio. Taken together, the centroid shift, CIE1931 *x*-coordinate, and $LIR_3$ provide three independent pressure-dependent observables, enabling spectroscopic, chromaticity-based, and ratiometric readout within the same phosphor. This multimodal sensing capability is particularly attractive from an application perspective, as it allows the pressure response to be evaluated using complementary optical parameters with different sensitivity ranges and readout characteristics. Considering the combination of high pressure sensitivity, broad useful pressure range, pronounced pressure-induced color evolution, multimodal optical readout, and reduced temperature cross-sensitivity, $KGaGeO_4:Bi^{3+}$,0.1%$Eu^{3+}$ emerges as a highly promising candidate for luminescence-based pressure sensing. The maximum sensitivity obtained for $KGaGeO_4:Bi^{3+}$,0.1%$Eu^{3+}$ is particularly high compared with previously reported ratiometric luminescence manometers operating in the visible spectral range below 700 nm. A comparison with representative literature examples is summarized in Table 2, below.

**Table 2.** Comparison of luminescence ratiometric (*LIR*-based) manometers reported in the literature operating in the spectral range below 700 nm.

| ***Luminescent manometer*** | ***$S_{R(p)}$ (% $GPa^{-1}$)*** | ***Reference*** |
|---|---|---|
| $La_6Sr_4(SiO_4)_6F_2:Ce^{3+}$ | 425 | 38 |
| $Eu(bpyO_2)_4(PF_6)_3$ | 120.7 | 39 |
| **$KGaGeO_4:Bi^{3+}, Eu^{3+}$** | **54.47** | **This work** |
| $NaYF_4:Er^{3+},Yb^{3+}$@$NaYF_4$ | 40 | 40 |
| $Sr_4Al_{14}O_{25}:Mn^{4+}$ | 24.8 | 41 |
| $K_2Ge_4O_9:Mn^{4+}$ | 21.7 | 42 |
| $CaLu^F:Er^{3+},Yb^{3+}$@CaLuF | 18.5 | 40 |
| $ZnAl_2O_4:Mn^{2+}$ | 18.34 | 43 |
| $SrLuF:Er^{3+},Yb^{3+}$@SrLuF | 14 | 40 |
| $SrB_4O_7:Eu^{2+}/Sm^{2+}$ | 13.8 | 44 |
| $BaLuF:Er^{3+},Yb^{3+}$@BaLuF | 11.1 | 40 |
| $Li_4SrCa(SiO_4)_2:Eu^{2+}$ | 10 | 45 |
| $NaErF_4:Tm^{3+}$@$NaLuF_4$ | 3.86 | 40 |
| $LaPO_4:Yb^{3+}/Tm^{3+}$ | 3.8 | 46 |
| $NaBiF_4:Yb^{3+}/Er^{3+}$ | 2.09 | 47 |
| $NaErF_4:Tm^{3+}$@$NaGdF_4$ | 1.82 | 40 |
| $NaErF_4:Tm^{3+}$@$NaYF_4$ | 1.37 | 40 |

For practical application, however, the performance of a luminescence pressure sensor cannot be assessed solely on the basis of its pressure sensitivity. Temperature cross-sensitivity is equally important, since variations in temperature may alter the same spectroscopic parameter used for pressure determination and consequently introduce an additional uncertainty in the pressure readout. Therefore, for the LIR-based ratiometric mode, the analogous temperature dependence of $LIR_3$ was evaluated and the corresponding temperature sensitivity was determined based on emission spectra as a function of temperature (Figure S10), as presented in Figure S11. To quantify the relative influence of pressure and temperature on the $LIR_3$ response, the Thermal Invariability Manometric Factor *TIMF*[48], was calculated using the pressure sensitivity and the temperature sensitivity determined at 303 K according to Eq. 10:

$$TIMF = \frac{S_{R(p)}}{S_{R(T)}(303K)} \quad \textbf{(11)}$$

where $S_{R(p)}$ represents the pressure sensitivity at a given pressure and $S_{R(T)}(303\ K)$ is the relative temperature sensitivity determined at 303 K. The maximum *TIMF* value of 114.9 K GPa$^{-1}$ was obtained, indicating that the $LIR_3$ response is governed predominantly by pressure rather than by temperature. This relatively weak temperature interference is particularly advantageous for practical pressure sensing under conditions where temperature fluctuations cannot be fully eliminated.

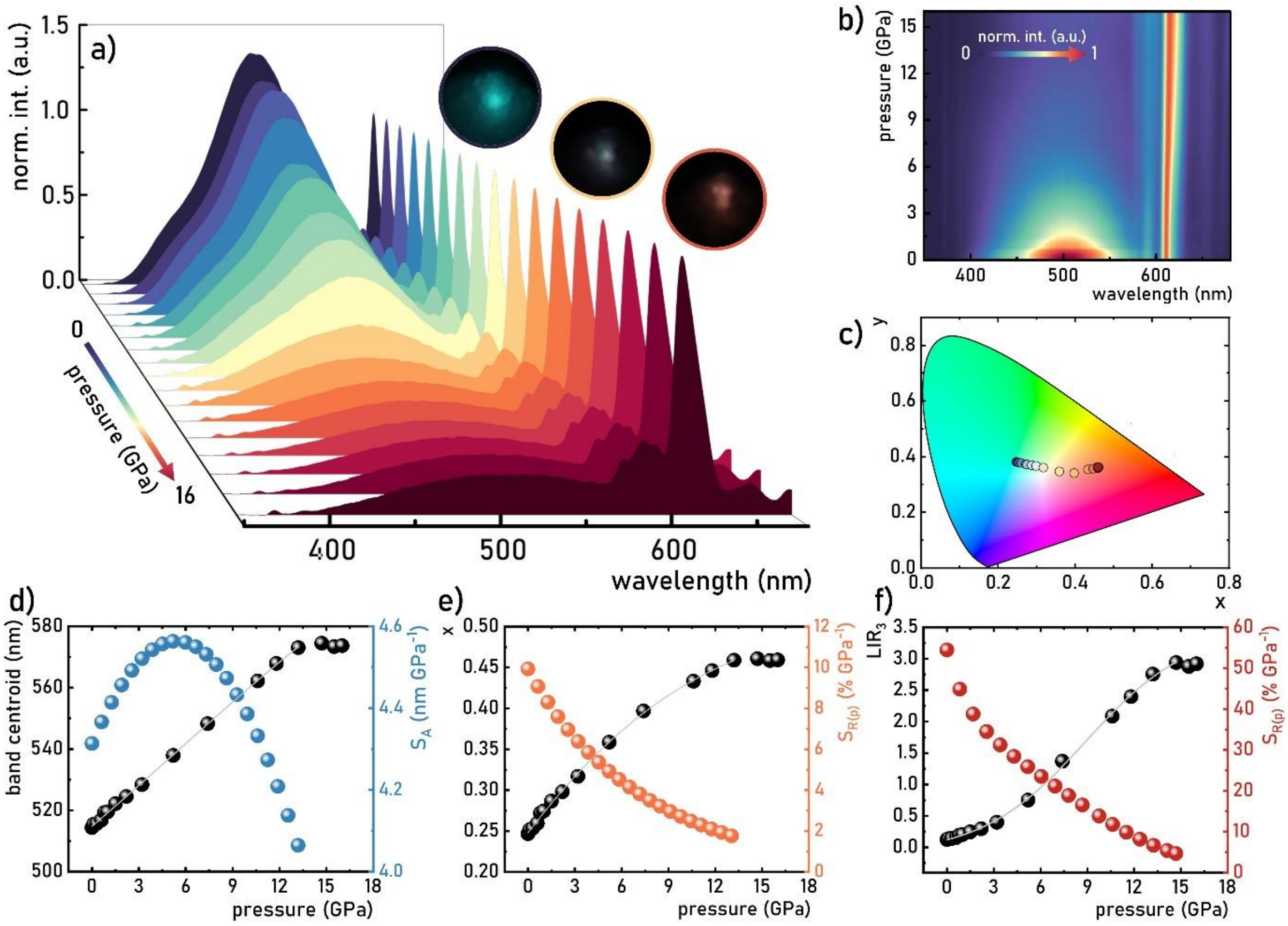

**Figure 7.** Pressure-dependent normalized emission spectra ($\lambda_{exc}$ = 355 nm), accompanied by photographs of the luminescence recorded in the DAC at 0, 5.2, and 14.7 GPa - a), pressure-dependent 2D emission map - b), evolution of the CIE1931 chromaticity coordinates with pressure - c), pressure dependence of the emission-band centroid - d), CIE1931 x coordinate - e), and LIR - f), together with the corresponding sensitivities (relative or absolute) as a function of pressure for $KGaGeO_4$:1%$Bi^{3+}$, 0.1%$Eu^{3+}$.

Considering the advantages of integrating luminescence imaging with multivariate data analysis, pressure-dependent luminescence images of $KGaGeO_4$:1%$Bi^{3+}$,0.1%$Eu^{3+}$ were recorded in the DAC (Figure 8b) and subjected to quantitative image analysis, following the workflow schematically illustrated in Figure 8a. Analysis of the individual RGB channels revealed distinct pressure-dependent responses (Figures 8c and S12). The R-channel intensity generally increases with pressure, whereas the G and B channels decrease and exhibit very similar trends. The markedly different pressure response of R compared with G and B makes ratios combining the R channel with either G or B particularly suitable as pressure-sensitive parameters. In contrast, the G/B ratio remains nearly unchanged because both channels respond to pressure in a comparable manner. Therefore, R/G and R/B were selected for further analysis. R/G exhibits a positive quadratic dependence over the investigated pressure range, whereas R/B increases monotonically up to 14.7 GPa. Accordingly, a phenomenological quadratic function was used to describe the R/G pressure dependence, while a linear calibration was applied to R/B up to 14.7 GPa (Figures 8d and 8e). The maximum $S_R$ values were obtained at ambient pressure and reached 169% GPa$^{-1}$ for R/G and 91% GPa$^{-1}$ for R/B. Within the 0.0001-7.4 GPa range, both ratiometric parameters exhibit approximately linear pressure dependences, enabling their simultaneous implementation in a MLR model. Here, as in previous thermometric studies, R/G and R/B were used as predictors and pressure as the target variable, as in Eq. 7. The standardized regression coefficients indicate that R/G provides the dominant contribution to the model, accounting for approximately 88%, whereas R/B contributes the remaining 12% (inset of Figure 8f and Table S2). Importantly, the simultaneous use of both predictors markedly enhanced the manometric performance, increasing the maximum relative sensitivity from 169% GPa$^{-1}$ for the best individual RGB ratio to 721% GPa$^{-1}$ for the MLR model, corresponding to an approximately 4.3-fold improvement. This enhancement demonstrates that the two ratios provide complementary pressure-dependent information that can be exploited more effectively when combined within a multivariate model. The good agreement between experimentally applied and MLR-predicted pressures confirms the reliability of the proposed approach (Figure 8g). To evaluate the cross-sensitivity of the RGB readout to temperature, the R, G and B channels were extracted (Figure S13) and the R/G and R/B ratios were analyzed as a function of temperature. Both parameters generally decrease upon heating from 93 to 673 K, although with different slopes. Importantly, within the intermediate temperature range of 313-513 K, both ratios remain comparatively stable, with variations not exceeding approximately 10%. This limited temperature-induced variation is also evident from the luminescence images recorded at 93, 373, and 673 K, for which the observed emission color remains predominantly within the blue region throughout the investigated range. Owing to the non-linear temperature dependences of the RGB ratios (Figure 8h), an analogous MLR calibration was not applied for thermal response. Instead, their $S_R$ were determined according to Eq. 9, yielding maximum values of 0.95% $K^{-1}$ for R/B and 0.68% $K^{-1}$ for R/G (Figure 8h). These values indicate that R/G is less affected by temperature variations, resulting in greater pressure-temperature discrimination and making it more suitable for reliable pressure sensing under non-isothermal conditions.

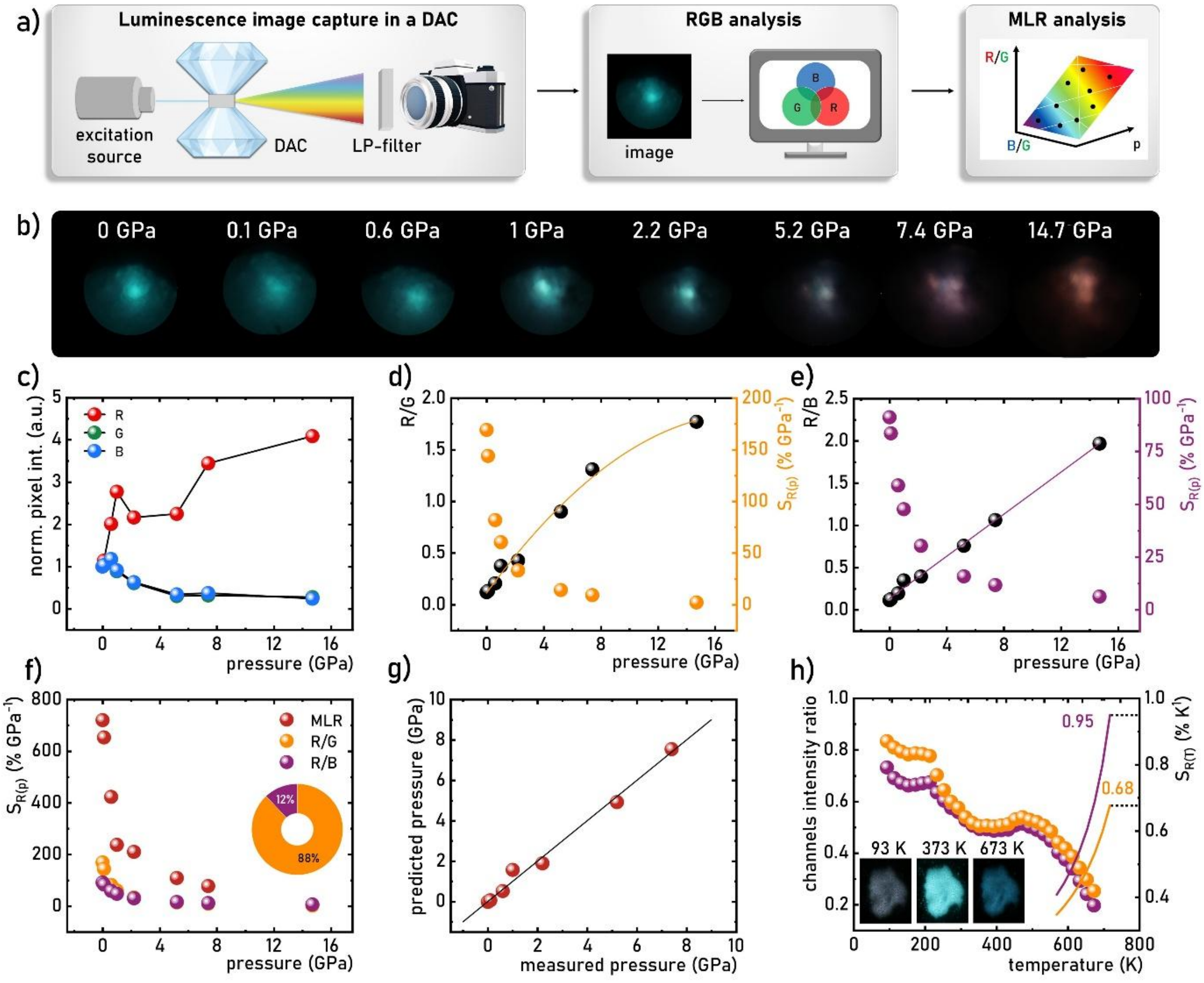

**Figure 8.** Schematic representation of the DAC-based luminescence imaging setup and the workflow used for RGB pressure analysis - a), pressure-dependent luminescence images recorded from ambient pressure to 14.7 GPa for $KGaGeO_4$:1%$Bi^{3+}$, 0.1%$Eu^{3+}$ - b), and normalized intensities of the blue (B), red (R) - and green (G)

channels extracted from the luminescence images -c). Pressure dependence of the R/G -d) and R/B -e) color-channel intensity ratios together with the corresponding relative sensitivities, $S_R$. Comparison of the $S_R$ values obtained using the MLR approach with those determined from the R/G and R/B ratios, with an inset doughnut chart showing the relative contributions of the R/G and R/B predictors to the MLR model; the values correspond to the standardized $\beta$-weights - f). Comparison between the pressure predicted by the MLR model and the experimentally applied pressure - g). Temperature dependence of the R/G and R/B channel intensity ratios together with the corresponding $S_R$, with inset luminescence images recorded at selected temperatures -h).

## Conclusions

This work establishes $KGaGeO_4:Bi^{3+},Eu^{3+}$ as a versatile and composition-tunable luminescent platform for visual, spectroscopic, and imaging-based sensing of both temperature and pressure. Importantly, variation of the $Eu^{3+}$ concentration enables the same host-dopant system to be selectively optimized toward either thermometric or manometric operation. This multifunctionality originates from the coexistence of spectroscopically distinct $Bi^{3+}$ centers and $Eu^{3+}$ emission, which exhibit markedly different responses to temperature and pressure and therefore generate strong, composition-dependent spectral redistribution.

For luminescence thermometry, $KGaGeO_4$:1%$Bi^{3+}$,5%$Eu^{3+}$ was identified as the optimal composition. Its pronounced temperature-induced redistribution between $Bi^{3+}$- and $Eu^{3+}$-related emission enabled several independent thermometric readout modes, including *LIR* and CIE chromaticity coordinates. The ratiometric approach provided a maximum relative sensitivity of 1.55% $K^{-1}$ at 299 K, with $S_R$ remaining above 1% $K^{-1}$ over the broad 213-393 K range. The strong temperature-dependent color evolution was additionally exploited for RGB-based sensing. By applying multiple linear regression (MLR), information from several color channels could be analyzed simultaneously, increasing the temperature sensitivity from 1 to 8.8 % $K^{-1}$. The practical potential of this approach was further confirmed through a proof-of-concept temperature-imaging experiment, demonstrating the possibility of spatially resolved thermal readout using luminescence color information.

In contrast, $KGaGeO_4$:1%$Bi^{3+}$,0.1%$Eu^{3+}$ was found to be the most promising composition for luminescence manometry because of its strong pressure response combined with reduced temperature interference. Compression induced a pronounced redistribution between the pressure-sensitive $Bi^{3+}$ emission and the comparatively pressure-stable $Eu^{3+}$ emission, resulting in a clearly visible color change from turquoise at ambient conditions through white to red at 16 GPa. Three complementary pressure readout modes based on the band centroid, CIE x-coordinate, and *LIR* were demonstrated, providing maximum sensitivities of 4.56 nm $GPa^{-1}$, 9.93% $GPa^{-1}$, and 54.47% $GPa^{-1}$, respectively. The *TIMF* analysis further confirmed the low temperature cross-sensitivity of the LIR-based mode, which is particularly important for reliable pressure determination under non-isothermal conditions. The pressure-dependent color response was also successfully translated into an RGB-based imaging strategy. As in the thermometric mode, the implementation of MLR enabled simultaneous utilization of the information encoded in multiple color channels and improved the pressure-readout performance from 169 % $GPa^{-1}$ to 721 % $GPa^{-1}$. This demonstrates that multivariate analysis can significantly enhance the sensing information extracted from luminescence images compared with conventional single-parameter RGB approaches. To the best of our knowledge, this represents the first demonstration of MLR-assisted RGB analysis for quantitative luminescence pressure sensing, establishing a new route toward simple and highly sensitive pressure readout directly from luminescence images.

Overall, the key outcome of this work is not only the demonstration of a dual temperature- and pressure-responsive phosphor, but the development of a composition-engineered, multimodal sensing concept in which the same $KGaGeO_4$:$Bi^{3+}$,$Eu^{3+}$ platform can be deliberately directed toward different sensing functions. The combination of concentration-controlled spectral response, multiple spectroscopic readout modes, pronounced visible color changes, RGB imaging, and MLR-enhanced analysis provides a powerful strategy for the design of next-generation luminescent sensors with high sensitivity, complementary readout capabilities, and strong potential for practical optical imaging applications.

**Conflict of Interest**

The authors declare no competing financial or commercial interests.

**Acknowledgements**

This work was supported by the Foundation for Polish Science under the First Team FENG.02.02-IP.05-0018/23 project with funds from the second Priority of the Program European Funds for Modern Economy 2021-2027 (FENG). Maja Szymczak gratefully acknowledges the support of the Foundation for Polish Science through the START program. This work is supported by the Fundamental and Interdisciplinary Disciplines Breakthrough Plan of the Ministry of Education of China (JYB2025XDXM403). Authors would like to acknowledge dr Damian Szymanski for SEM and EDS analyses.